# Further Argument Against the Motion of the Earth, Based on Telescopic Observations of the Stars: An English rendition of chapter 30, book 9, section 4, pages 460-463 of the *Almagestum Novum* Volume II of G. B. Riccioli.

Christopher M. Graney
Jefferson Community & Technical College
1000 Community College Drive
Louisville, KY 40272
christopher.graney@kctcs.edu

The Italian Jesuit astronomer Giovanni Battista (Giambattista) Riccioli constructed a powerful, thoroughly scientific argument in favor of geocentrism – an argument based on telescopic observations of stars. This paper contains a rendition of a chapter from Riccioli's *Almagestum Novum* in which he lays out this argument. Riccioli opposed Galileo's heliocentrism, but like Galileo, relied on telescopic observations to make his case. Riccioli's observations and argument were both valid, lacking only in that the functioning of the telescope was not fully understood at the time.

Geocentrists who opposed the "Copernican Revolution" are typically portrayed as churchmen or Aristotelian scholars disinterested in, not persuaded by, or perhaps even hostile to knowledge gathered through scientific inquiry by way of the newly invented telescope. Albert Einstein, in writing on Galileo Galilei's *Dialogue Concerning the Two Chief World Systems*, provides an example of such an image in the way he contrasts Galileo with those who opposed him. To Einstein, Galileo is –

> …a representative of rational thinking against the host of those who, relying on the ignorance of the people and the indolence of teachers in priest's and scholar's garb, maintain and defend their positions of authority.[1]

– and Galileo was able to –

> …overcome the anthropocentric and mythical thinking of his contemporaries and lead them back to an objective and causal attitude toward the cosmos….[2]

To Einstein, geocentrists relied on authority, opposed rational and objective thinking concerning nature, were grounded in anthropocentric and mythical thinking, and were ignorant and lazy, too. Yet the Italian Jesuit astronomer Giovanni Battista Riccioli (1598-1671; also Giambattista), supported geocentrism with a powerful, thoroughly scientific argument based on telescopic observation, rational and objective thinking concerning nature, and hard work. Telescopic observations of stars at the time provided a strong scientific case for geocentrism, something that has been overlooked in the history of science until recently.[3]

What follows is a rendition in English of chapter 30, book 9, section 4, pages 460-463 of *Almagestum Novum* Volume II.[4] It is not a translation – I relegate most mathematical calculations to footnotes, and those calculation I present in modern form; I do away with Riccioli's 150-word sentences that contain multiple colons; I bring material he provides in tabular form into his narrative text. This is my attempt to present Riccioli's argument in a manner accessible to the informed modern reader who is not a specialist in astronomy, leaving out nothing Riccioli says and adding nothing to what Riccioli says.

## CHAPTER XXX

*An argument is put forward against the annual motion of the Earth based on the excessive sizes of stars. A discussion follows concerning whether such excessive size is more incredible than the speed those stars must have in a world system where the Earth is at rest.*

1. The unreasonably huge size of the stars (whether compared to the Earth, the sun, or the annual orb[5]), which follows from their distances in the Copernican hypothesis, is clear. Tycho, Longomontanus,[6] Scheiner,[7] Claramontius[8] (each discussed below), and some others who have noted this, have been dismissed by Copernicans because of the excessive apparent diameters of stars that they claimed.

Tycho says in a letter to Rothmann of 1589 November 24 (page 167) that even third magnitude stars,[9] which have diameters of only a minute of arc, will be as big as the annual orb – 2284 Earth radii[10] – based on their being at a distance of 7,850,000 Earth radii. Tycho asks what then of first magnitude stars, which can have diameters of two or even three minutes?

2. Longomontanus investigates the magnitude of the stars, just as we [Grimaldi and I] have done (see *Almagestum Novum* [abbreviated as *A.N.* from here on] book 6, chapter 9, number 9; see Longomontanus book 1, chapter 1, the first page under the heading “Theoricorum”):[11]

Refer to Figure 1. Let ABC be an arc of the circumference of the highest heaven – the sphere of stars – whose center is G. Let DHEF be some fixed star with center B. Lines GD and GE may be drawn from G tangent to the globe of the star at D and E. The star’s radius is BD or BE and the star subtends an angle DGE which is its apparent diameter. Triangle DBG is a right triangle, so the star's true diameter (twice BD), its distance BG, and its apparent diameter are all related through basic triangle trigonometry. If the star's true diameter is known in terms of the diameter of the Earth, then cubing the star's diameter will indicate how many times the volume of the Earth is contained within the volume of the star.

Longomontanus, by reason of the Copernican hypothesis, determines BG to be 7,906,818 Earth radii. He assumes the apparent radius of a first magnitude star (angle BGD) to be one minute.[12] From these numbers and basic trigonometry he calculates the true diameter of a first magnitude star according to the Copernican hypothesis to be 2300 Earth diameters – a very large number. The star must be 12,167,000,000 times more voluminous than Earth. Since the sun is 140 times more voluminous than Earth, the star is in turn 86,907,143 times more voluminous than the sun. And since the star, at 2300 Earth diameters, is approximately double the diameter Copernicus gives for the annual orb (1142 Earth diameters), the star is roughly 8 times as voluminous as the entire annual orb.[13]

Hence Longomontanus concludes that based on the following two reasons – the vast interval, the great discontinuity, between the most distant of the planets and the sphere of the stars; and the incredible size of the stars that follows from that vast distance – we should get rid of the Copernican hypothesis. He says these absurdities heedlessly destroy the symmetry of the universe. He further notes that we should particularly get rid of two parts of that hypothesis, those being the annual motion of the Earth, and the libration of the Earth's poles.[14]

3. Scheiner in *Disquisitionibus Mathematicis*[15] (page 26) derives the distance to the stars according to the Copernican hypothesis to be 13,133,376 Earth radii. He assumes the sun's distance to be 1208 Earth radii, and its true diameter to be 11 Earth radii. He takes the apparent diameter of the sun at apogee to be 30', and of a star of first magnitude to be 2'.[16] Then he calculates the true diameter a first magnitude star to be 7972 Earth radii, or 3.3 times the diameter of the annual orb.[17] Likewise he determines the true diameter of a third magnitude star (apparent diameter 1') to be 2416 Earth radii or 1.6 times the diameter of the annual orb. Thus a first magnitude star is almost 32 times more voluminous than the annual orb.[18]

Therefore on page 28 Scheiner concludes that, as the Copernican movement of the Earth requires the largest star to be thirtyfold the entire space enclosed by the sun,[19] and requires even a little star to be very large, he will admit no such movement of the Earth.

4. By the same method Claramontius in defending his *Antitycho* (part 3 chapter 12 and 13) calculates that, according to the Copernican hypothesis, a first magnitude fixed star is 1,349,232,625 times more voluminous than Earth and 8,079,237 times more voluminous than the sun. Moreover, Lansbergius in *Uranometria* (book 3 element 20)[20] grants a first magnitude star a diameter of 1', grants stars a distance of 41,958,000[21] Earth radii, and grants the annual orb a diameter of 1498.5 Earth diameters. Calculating via the method shown above (Figure 1), he establishes a first magnitude star to be 20,053 times more voluminous than Earth.[22] Lansbergius says that sixth magnitude stars have an apparent radius of 2.5", corresponding to being 11.5 times more voluminous than the annual orb.[23]

For more information concerning stars, see what I have reported in *A.N.* book 6 chapter 9.

5. Still, Hortensius, an astronomer who is a Copernican, lists first magnitude stars as being only 8" in diameter, and sixth magnitude stars as being only 2".[24] From these values and from the distance he gives for the fixed stars of 10,312,227 Earth radii (assuming a radius of the annual orb of 1498.5 Earth radii),[25] he calculates a first magnitude star to be 422 times *less* voluminous than the annual orb, and a sixth[26] magnitude star to be 27,826 times *less* voluminous.

But Galileo makes the stars smaller still. In the *Dialogue Concerning the Two Chief World Systems* (page 265 in the Latin version;[27] page 352 in the Italian version) he reproves those who "set themselves to calculating...and deduce that in Copernicus's doctrine one must admit that a fixed star is much larger than the orbit of the earth".[28] He shows how to measure the diameter of stars (page 267, Latin), and he asserts that the diameter of a first magnitude star does not exceed 5" (page 268, Latin).[29] This yields a far smaller star; see *A.N.* book 6, chapter 9, number 3.

6. P. Francisco Maria Grimaldi and I have developed a method for reliably measuring the apparent diameters of the stars [by means of a telescope] (see *A.N.* book 6 chapter 9 number 6; book 7 section 6 chapters 9-11). By this method I have measured the most prominent of stars, Sirius, to have an apparent diameter of 18". Likewise, I have determined Alcor, in the tail of Ursa Major and one of the smallest[30] stars capable of being seen by the naked eye, to have an apparent diameter of 4" 24'" [see Figure 2].

Using these [telescopically measured] apparent diameters, and opinions of various astronomers concerning the distances of the fixed stars, I have deduced various possible values for their true sizes. Those values can be found in three tables set forth in *A.N.* chapter 11, the last two of which depend on Copernican distances. Those two tables are [partially] reproduced here as Tables 1 and 3, and they show how many times more voluminous than Earth are Sirius and Alcor. Sirius and Alcor represent the largest and smallest of the fixed stars visible to the naked eye, so all stars fall somewhere in between these two. The accompanying Tables 2 and 4 show how voluminous Sirius and Alcor each are in comparison to the annual orb.[31]

7. These tables show the following: If the Earth circles the sun, then the stars must be exceedingly, vastly, hugely voluminous. They must be huge compared to the Earth, huge compared to the sun, even huge compared to the annual orb. It is not merely the most prominent star that is huge (as seen in Table 4, Sirius is at least 44 times more voluminous than the annual orb), but also even the least of the stars (Alcor is at least 4 times more voluminous than the annual orb). But this is absurd and incredible.

So the tables show how the stars compare to the other bodies in the Universe, and what a sphere of giant stars does to the symmetry of the Universe. Rothman himself, even though he was a Copernican, in the end conceded to Tycho that if the Copernican hypothesis was true then most stars would surpass the entire annual orb. Rothmann admitted that the Copernican hypothesis logically led to the stars being so great in size as to be beyond all belief, to be absurd, to be asymmetric, and to be unexplainable except perhaps by means of Divine Omnipotence.

As for Galileo and Hortensius, as discussed in number 5 above – obviously Tycho, Longomontanus, Scheiner, and some others overestimated the diameter of the stars; but Galileo and Hortensius have underestimated them. Table 1 shows the sizes of stars calculated[32] using our observations of Sirius and Alcor and using Galileo's and Hortensius's assumptions about the size of the annual orb and the distance of the stars. Table 2 shows that under these assumptions the stars are large, although perhaps not excessively so if the standard of comparison is the annual orb.

But Table 3 shows the sizes of stars calculated using our observations of Sirius and Alcor, using Galileo's and Hortensius's assumptions about the size of the annual orb, and using the minimum distance of the stars required in order that the Earth's annual motion will create an annual parallax of no more than 10''. Table 4 shows how both stars are exceedingly large under these more valid calculations – Sirius being at least 44 times the volume of the annual orb, while Alcor [whose size falls in the ranges given by Hortensius and Galileo] is at least 4 times the annual orb's volume.[33]

However this is done, there is no avoiding the Copernican large size problem.[34]

A Copernican might then simply deny that this is a problem – the vastness of the stars might be claimed as evidence of Divine Magnificence. If the various bodies in the Universe are compared to one another, by any measure we find that the Sun is far larger than the Earth – perhaps 140 or 160 times more voluminous, perhaps 38,600 times if the distance to the sun is as we calculate from our Lunar work, perhaps 262,144 times if the distance is what Vendelinus calculates (see *A.N.* book 3, chapter 11, problem 1). Thus the stars are larger than the sun, and the sun is larger than the Earth. In the end a Copernican will compare the vast sizes the stars must have in the Copernican hypothesis to the speed the stars must have in the Ptolemaic hypothesis, and declare the Copernican side to be much more credible.

None of the Copernican reasoning is satisfactory.

If God's intention was for the vastness of the stars to show His Greatness to mortals, he did well. He set them so far distant from us that they appear small – barely one hundredth the apparent diameter of the sun. He might have provided a few other pieces of information that would have allowed us to come into certain knowledge of this distance and size.

Yet he did not. In fact all astronomical phenomena [both those visible to the naked eye and those seen with the telescope] can be fully explained by way of a geocentric hypothesis [such as that of Tycho Brahe]. The Copernican hypothesis is not needed. Moreover, experiments involving the physics of gravity and projectiles clearly contradict that bogus hypothesis.

Furthermore, the sun being large compared to Earth seems quite reasonable. It clearly holds an important function in the universe. [The telescope has shown that] it is the source of light for all the planets [and the moon]. It governs the universe above the moon and below the stars.[35]

But such reasoning does not extend to explaining the little stars that supposedly are actually so much larger and so much more distant than the sun. There is no apparent reason for such vast bodies, completely removed from the planetary system. Sacred Scripture – Deuteronomy certainly – says the heavenly bodies are created to serve us; they are not to be worshipped because they are to serve us, not master us.[36]

Now let us address the question of whether the Copernican vast sizes of unmoving stars is preferable to the well-known Ptolemaic speed of the stars.

*A discussion of whether the vast sizes of the stars in the Copernican hypothesis is more credible than the speed of the stars in the Ptolemaic hypothesis.*

8. Regarding the incredible (or not) velocity of the stars in the Ptolemaic hypothesis – see *A.N.* chapter 6, numbers 1-7 where we have enumerated the opinions and calculations of Copernicans in opposition to this speed; see numbers 7-8 where we have refuted their arguments, which are so much equivocation and sophistical tricks. The unprejudiced person should read these and pass judgment – certainly against the Copernicans.

Some points to keep in mind:

Consider a rotating wheel. All points on the wheel move with a common motion. Points further from the center of the wheel move more swiftly than those closer to the center. The same is true for a rotating sphere. So, in the case of the stars, if we consider them to all be circling the Earth with common motion, the more distant we assume them to be, the faster they must be moving.[37]

We may consider the speed of the fixed stars to be excessive because we express that speed in our units – in human miles which are constructed based on human strides. The speed of a walking Giant would seem to us to be great, and as a Giant would cover an ant mile in a single stride, his speed would seem exceedingly great to an ant. We must be careful to be mindful of our standards of comparison when discussing star sizes and speeds as regards comparing Ptolemy against Copernicus.

Finally, as concerns this issue of comparison, I have no quarrel here with the semi-Copernicans who say the Earth's motion is limited to daily rotation.[38] These include Longomontanus, Origanus, Argolus, and others. They see the absurd Copernican size of the stars. They transfer the issue of swiftness from the stars down to the Earth.

Moving along with our discussion, the Ptolemaic velocity of the stars depends on the circumference of the heavenly sphere of stars, which revolves about the Earth in a 24 hour period. In discussing the issue of the swiftness of stars, the Copernicans are comparing that circumference to the circumference of the Earth. Therefore, to make a fair comparison, the globes of the stars should also be compared against the globe of the Earth. Let us do just that, and see what is more extreme, more disproportionate – the circumference about which the stars must travel in a geocentric hypothesis, or the star size that is unavoidable in a heliocentric hypothesis.

First the geocentric hypothesis: According to Ptolemy, the sphere of stars measures 20,220 Earth diameters. According to Alfeganus the value is 40,440 (see Theorem 5, book 11 of the collection of Pope Alexander). Our opinion is that it is more like 100,000 Earth diameters, or even 210,000 (see *A.N.* book 6, chapter 7, number 6 and Table 1). Now the Copernican hypothesis: Consulting Table 1 and Table 3, we see that not only Sirius, the greatest of stars, but even Alcor, are more voluminous than Earth by at least tens of millions of times, perhaps as much as trillions of times.

If the Copernicans would rather we compare the fixed stars to the annual orb, so the numbers seem more reasonable, we shall not object – provided that, in the interest of being fair and balanced, we compare the geocentric circumferences with the annual orb as well. So in table 4 we see that Sirius is at least 44 times more voluminous than the annual orb. The geocentric circumference of the starry sphere is 16.71 times the circumference of the annual orb if we use Ptolemy's figures, 13.70 or 28.77 times if we use ours. Thus even if the standard of comparison is the annual orb, the Copernican hypothesis shows the greater disproportionality.[39]

All things being equal, a Universe composed of bodies of moderate sizes, with swift movement by some of them, is more credible, and shows greater commensurability, than a Universe composed of bodies of immoderate sizes, with slower movement by some and no movement by others.

And lastly there is Sacred Scripture. Through the Psalmist God himself upholds the speed of the Sun:

> He hath set his tabernacle in the sun: and he, as a bridegroom coming out of his bride chamber, Hath rejoiced as a giant to run the way: His going out is from the end of heaven, And his circuit even to the end thereof: and there is no one that can hide himself from his heat.[40]

Nowhere does Scripture mention stars having sizes that vastly exceed the Sun and the Earth – it only mentions their innumerable multitude. Scripture upholds one Sun – the Giant in the abovementioned Psalm, and also:

> The firmament on high is his beauty, the beauty of heaven with its glorious shew. The sun when he appeareth shewing forth at his rising, an admirable instrument, the

> work of the most High. At noon he burneth the earth, and who can abide his burning heat?[41]

Apparently to the Copernican hearing the Giant is a Pygmy, and the admirable instrument is the contemptible little tool, versus the vast stars.

And so to conclude:

*More credible is the daily swiftness of the fixed stars in the hypothesis of a resting Earth; than the insanely large masses of the same in the hypothesis of the Earth wandering through the annual orb.*[42]

ACKNOWLEDGEMENT

I thank my wife, Christina Graney. Without her help in translating Riccioli's Latin this paper would not exist.

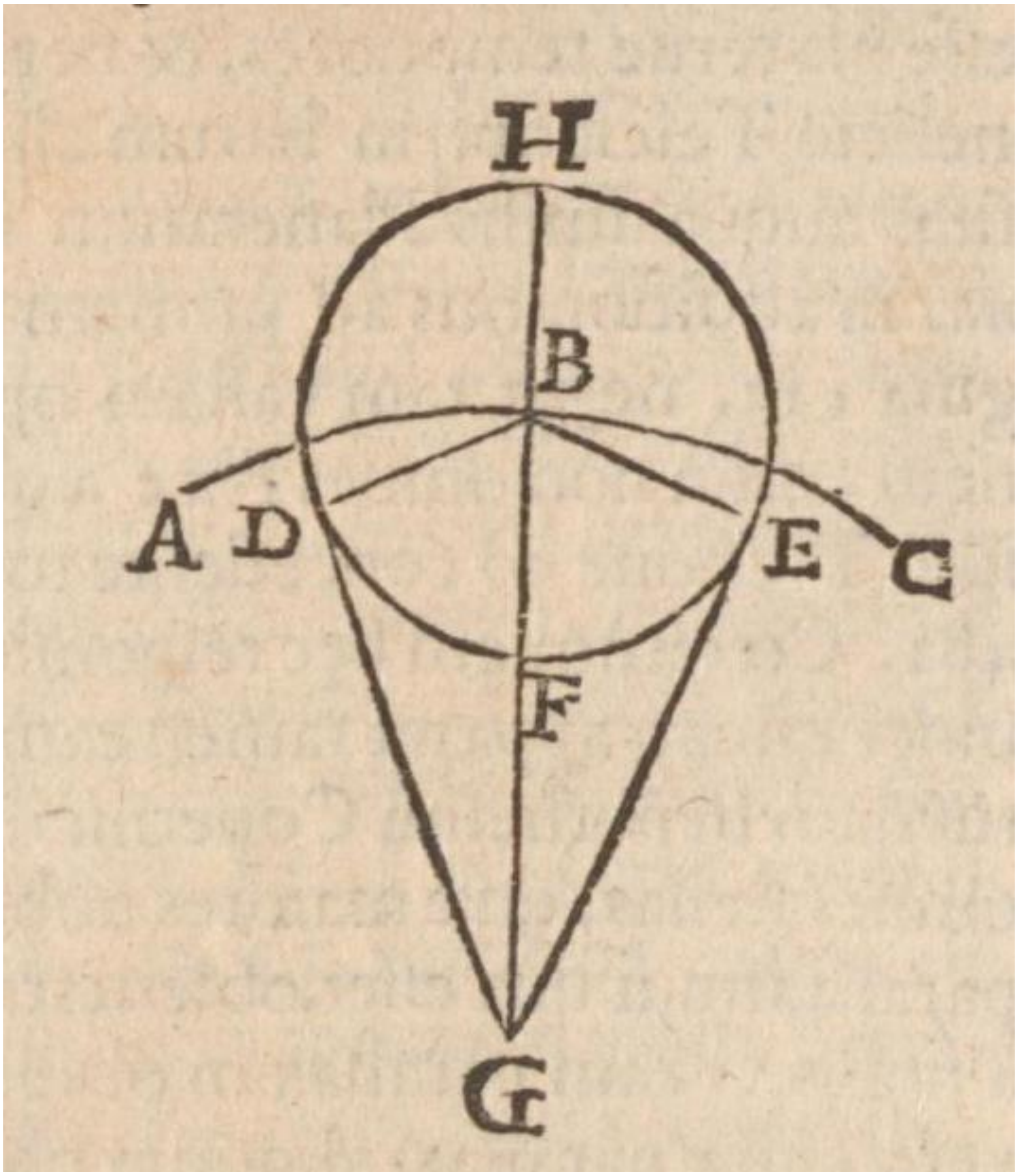

FIGURE 1. Diagram from the *Almagestum Novum* showing distances and angles in Riccioli's discussion of star sizes.

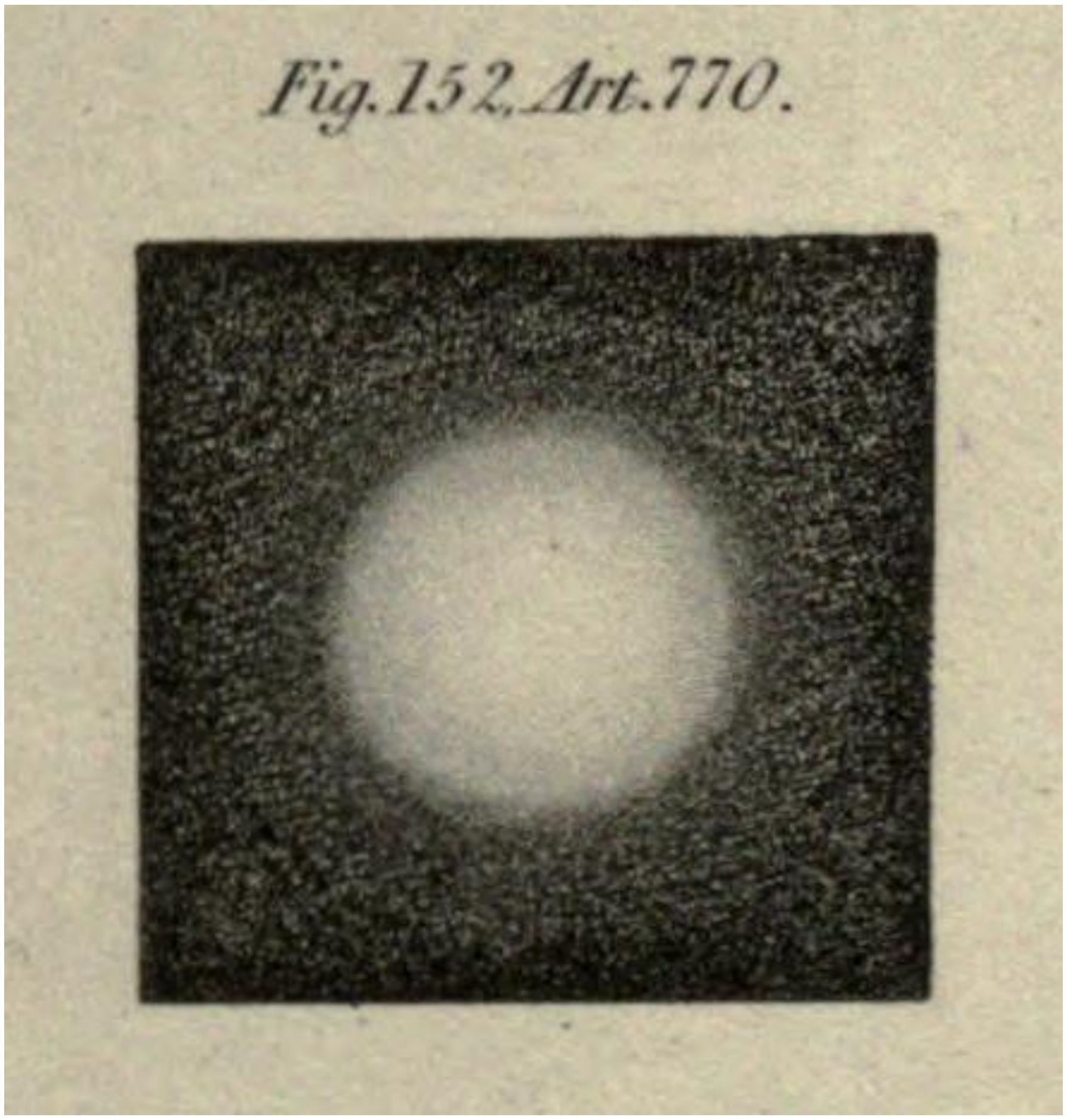

FIGURE 2.  The appearance of a star as seen through a telescope of very small aperture. This image is from John Herschel's article "Light" in *Treatises on Physical Astronomy, Light and Sound Contributed to The Encyclopaedia Metropolitana* (London, 1828).  That this globe-like image is an entirely spurious artifact of the telescope was not understood at the time of Galileo and Riccioli.  For details, see Christopher M. Graney and Timothy P. Grayson, "On the Telescopic Disks of Stars: A Review and Analysis of Stellar Observations from the Early Seventeenth through the Middle Nineteenth Centuries", *Annals of Science*, DOI: 10.1080/00033790.2010.507472 (iFirst on-line edition published 27 October 2010 – print edition in press).

I. TABVLA. Magnitudo Vera Fixarum Maximæ ideſt SIRII, & Minimæ nudo oculo viſibilium ideſt ALCOR; poſita diametro apparenti Sirij 18″. & Alcor 4″. 24‴. & Diſtantia Fixarum aſſerta à quatuor infraſcriptis Copernicanis.

| Ex Auctoribus | Fixarum Diſtantia in ſemidiam. Terræ | Semidiameter Orbis Annui Semid. Ter. | Continet ergo SIRII: Diameter Ter. Diam. | Continet ergo SIRII: Corpus Terram Vicibus | Continet verò ALCOR: Diameter Ter. Diam. | Continet verò ALCOR: Corpus Terram Vicibus |
|---|---|---|---|---|---|---|
| *Hortenſio* | 10 312 227 | 1498 ½ | 899 | 726 572 600 | 442 | 86 355 888 |
| *Galilæo* | 13 046 400 | 1208 | 1138 | 1 473 760 072 | 558 | 173 741 112 |
| *Lansbergio* | 41 958 000 | 1498 ½ | 3658 | 48 947 466 312 | 1796 | 5 793 206 336 |
| *Keplero* | 60 000 000 | 3469 | 5232 | 143 219 897 228 | 2568 | 16 933 994 432 |

| II. TABVLA. | ERGO SIRIVS | Vicibus | ET ALCOR | Vicibus |
|---|---|---|---|---|
| *Hortenſio* | Continetur in Orbe Annuo | ferè 5 | Continetur in Orbe Annuo | 39 ½ |
| *Galilæo* | Continetur in Orbe Annuo | ferè 1 2/11 | Continetur in Orbe Annuo | 10 |
| *Lansbergio* | Continet Orbem Annuum | ferè 145 | Continet Orbem Annuum | 1 12/17 |
| *Keplero* | Continet Orbem Annuum | ferè 3 1/3 | Continetur in Orbe Annuo | 2 9/17 |

III. TABVLA. Magnitudo Vera SIRII & ALCOR. Suppoſita earum Semidiametro Apparenti, Sirij 18″. & Alcor 4″. 24‴. & Diſtantia Fixarum tanta, vt Parallaxis Orbis Annui non excedat Secunda 10″. & ſuppoſita Semidiametro Orbis Annui ex infraſcriptis Auctoribus.

| Ex Auctoribus | Fixarum Diſtantia Semidiametrorum terreſtrium | Semidiameter Orbis Annui. Semid. ter. | Continet ergo SIRII: Diameter Diam. terr. | Continet ergo SIRII: Corpus verò Terram Vicibus | Continet autem ALCOR: Diameter Diam. terr. | Continet autem ALCOR: Corpus verò Terram Vicibus |
|---|---|---|---|---|---|---|
| *Copernico* | 47 439 800 | 1142 | 4170 | 71 677 713 000 | 1992 | 4 378 454 048 |
| *Herigonio* | 49 502 400 | 1200 | 4350 | 82 312 875 000 | 2068 | 8 844 058 432 |
| *Galilæo* | 49 832 416 | 1208 | 4380 | 88 427 672 000 | 2092 | 9 155 562 688 |
| *Bullialdo* | 60 227 920 | 1460 | 5300 | 148 877 000 000 | 2530 | 15 941 277 000 |
| *Lansbergio* | 61 616 122 | 1498 ½ | 5424 | 159 371 956 024 | 2588 | 17 333 761 472 |
| *Keplero* | 142 746 428 | 3469 | 12550 | 1 976 656 375 000 | 6000 | 216 000 000 000 |

| IV. TABVLA. | Continet Ergo Orbem Annuum SIRIVS Vicibus | ALCOR Vicib. |
|---|---|---|
| *Copernico* | 48 | 4 |
| *Herigonio* | 48 | 4 1/5 |
| *Galilæo* | 52 | 5 1/3 |
| *Bullialdo* | 44 | 4 2/3 |
| *Lansbergio* | 47 | 5 |
| *Keplero* | 46 | 5 |

TABLES 1-4 from the *Almagestum Novum*.

[1] Galileo Galilei, *Dialogue Concerning the Two Chief World Systems – translated and with revised notes by Stillman Drake; foreword by Albert Einstein; introduction by J. L. Heilbron* (New York, 2001), p. xxiii.

[2] *Dialogue*, p. xxiii.

[3] Christopher M. Graney, "The Telescope Against Copernicus: Star Observations by Riccioli Supporting a Geocentric Universe", *Journal for the History of Astronomy*, vol. 40 (2010); Christopher M. Graney, "Seeds of a Tychonic Revolution: Telescopic Observations of the Stars by Galileo Galilei and Simon Marius", *Physics in Perspective*, vol. 12 (2010).

[4] G. B. Riccioli, *Almagestum Novum* (Bologna, 1651) . <http://www.e-rara.ch/zut/content/pageview/140188 >
Riccioli does not make a different argument here than he does in the portions of his work included in Graney, "The Telescope Against Copernicus…". However, the manner in which he presents the argument differs, and the material may be of general interest to a scholar who wishes to study Riccioli.

[5] "comparentur cum … toto Orbe Annuo". The "annual orb" is the sphere defined by Earth's orbit around the sun in the heliocentric hypothesis. $17^{th}$-century astronomers understood it to be far larger than even the sun.

[6] Christen Sørensen Longomontanus (1562 – 1647), a Danish astronomer who served as an assistant to Tycho. He promoted Tycho's geo-heliocentric hypothesis, but attributed diurnal rotation to the Earth and not the heavens.

[7] Christoph Scheiner (1573-1650), an accomplished German Jesuit astronomer, perhaps most famously known for his disputes with Galileo over the nature of sunspots.

[8] Scipio Claramontius (1565-1652), also Chiaramonti. His *Anti-Tycho* is discussed in several places in Galileo's *Dialogue* (see *Dialogue* pp. 59, 65, 287, 313).

[9] Stars were rated by magnitude. The brightest stars visible to the naked eye were said to be of the first magnitude (first-rate stars). The faintest stars visible to the naked eye were of the sixth magnitude (sixth-rate stars). This system is still used by astronomers today, with some modifications.

[10] Riccioli cites Copernicus for the value for the radius of the annual orb – 1142 Earth radii (diameter 2248).

[11] Longomontanus's *Astronomica Danica* (Amsterdam, 1622) contains this work (page 18) under the heading "Theoricorum", although the diagram used is somewhat different.

[12] $^{1}/_{60}$ of a degree.

[13] Riccioli does not give the details of these calculations, but they are worth including as such calculations are central to Riccioli's discussion:
Referencing figure 1, the star's apparent diameter DGE according to Longomontanus is 1'. Thus angle DGB is 0.5'. Longomontanus gives the distance to the star BG to be 7,906,818 Earth radii.

By right triangle trigonometry distance DB is related to these values by {DB/BG = sin DGB}, or {DB/7,906,818 = sin (1/30)º}. Thus DB = 1150 and the true diameter of the star is 2300 Earth radii. {$2300^3$ = 12,167,001,843}, Riccioli rounds this to 12,167,000,000. {12,167,000,000/140 = 86,907,143}. {(2300/1142) ≈ 2}, and {$2^3$ = 8}.

[14] Longomontanus page 19. Libration – if the Earth is observed from the sun (the center of Earth's annual motion) in the Copernican system, the poles of the Earth rotate towards and away from the sun. This was the third motion of the Earth, after the diurnal rotation and the annual revolution. Tycho had remarked that the Copernican system

> …expertly and completely circumvents all that is superfluous or discordant in the system of Ptolemy. On no point does it offend the principle of mathematics. Yet it ascribes to the earth, that hulking, lazy body, unfit for motion, a motion as quick as that of the aethereal torches, and a triple motion at that [Owen Gingerich, *The Eye of Heaven: Ptolemy, Copernicus, Kepler* (New York, 1993), p. 181.].

In the *Dialogue* Simplicio complains that "the earth is made to move in three discordant and distractingly different ways" (page 312). We understand this today to be the root cause of the seasons, and it we understand it in terms of the physics of angular momentum. At the time, however, such physics had not been developed.

[15] Christophori Scheiner, *Disquisitiones Mathematicæ* (Ingolstadt, 1614).

[16] Units for measuring apparent sizes in the sky: ' – a minute of arc (1/60 of a degree); " – a second of arc (1/60 of a minute); "' – a third of arc (1/60 of a second).

[17] Riccioli provides the details of this calculation. First, calculate the true diameter the sun would have were it located on the stellar sphere but still had a visual diameter of 30':

(11/1208) = ($X$/13,133,376); so $X$=119,592 Earth radii.

Then use this to calculate the true diameter of a star with visual diameter of 2':

(30'/2') = (119,592/$D_{star}$); $D_{star}$ = 7972.

[18] This value should be 36, as $3.3^3 \approx 36$. Calculation errors are a problem for Riccioli, as shall be seen in further notes.

[19] The space enclosed by the orb of the sun in a geocentric hypothesis is the same as the annual orb in a heliocentric hypothesis. Scheiner is speaking from a geocentric point of view.

[20] Phillipi Lansbergi, *Uranometria*, *liber tres* (Middleburg, 1631), pp. 130-134. Philips Lansbergen (1561 – 1632) was a Dutch astronomer.

[21] The number 41,958,000 appears to be erroneous. The value Lansbergius gives is 280,000,000 (see *Uranometria*, p. 130). The other values Riccioli relates for Lansbergius are correct (*Uranometria*, p. 130-133).

[22] Lansbergius fully accepted enormous stars in the Copernican system on the basis of religious reasons – see Rienk Vermij, "Putting the Earth in Heaven: Philips Lansbergen, the Early Dutch Copernicans and the Mechanization of the World Picture," in *Mechanics and Cosmology in the*

*Medieval and Early Modern Period – Massimo Bucciantini, Michele Camerota, Sophie Roux, eds.* (Florence, 2007).

[23] Riccioli provides the details of these calculations. He says that Lansbergius calculates that a first magnitude star has a true diameter of 40,712 Earth diameters. As the value given for the annual orb is 1498.5, then $\{40{,}712^3 / 1498.5^3 = 20{,}053\}$. For the sixth magnitude star Riccioli gives a true diameter of 3388 Earths, but does not provide the volume calculations.

[24] M. Hortensius, *Dissertatio de Mercurio in Sole viso* (Lieden, 1633), 61-62. Hortensius – Martin (Maarten) van den Hove (1605 – 1639) – was a Dutch astronomer.

[25] Hortensius gives the distance to the fixed stars as 68,754,937 units, where the radius of the annual orb is 10,000 units. The value of 1498.5 does not appear in the pages of *Mercurius* cited here, so Riccioli has reworked Hortensius's calculations.

[26] The *A.N.* says "tertiæ" but the number stated, 27,826, is Hortensius's value for sixth magnitude. The number 422 is Hortensius's value for first magnitude.

[27] *Systema Cosmicum*, Lyon, 1641

[28] This quote from the Drake translation of the *Dialogue*, p. 417. Riccioli does not use a quote directly but echoes the language from the *Dialogue*.

[29] For the discussion described here, see *Dialogue*, pp. 417-423.

[30] Today astronomers would say "faintest," but at this time astronomers thought of stars in terms of size.

[31] Riccioli states that he did this calculation by dividing the number of Earth volumes contained within the star by the number of Earth volumes contained within the annual orb, and points the reader to *A.N.* chapter 29, number 15, column 1 of table 3.

[32] Riccioli does not include the calculations for Tables 1 and 2. The following example is for the reader who is interested in such details, which do help to illustrate certain strengths and weaknesses of Riccioli's argument:

As an example, in the second row of Table 1, Riccioli states that Galileo attributes a distance to the fixed stars of 13,046,400 Earth radii (2nd column), and a radius of the annual orb of 1208 Earth radii (3rd column). Thus the distance to the stars is 10,800 times the radius of the annual orb (see *Dialogue*, pp. 424, 571, keeping in mind that Galileo provided greatly varying estimates of stellar distances, from a couple of thousand annual orb radii or less – *Dialogue* p.417 – to 27,000 annual orb radii – *Dialogue* p. 424).

A circle with radius 13,046,400 has circumference of $\{2\cdot\pi\cdot 13{,}046{,}400 = 81{,}972{,}949\}$. Sirius has diameter 18", or (18/3600)°. As there are 360° in a circle, this is 18/(3600·360) or 18/1,296,000 of a circumference. $\{(18/1{,}296{,}000)\cdot 81{,}972{,}949 = 1139\}$, so the true diameter of Sirius is 1139 Earth radii. Riccioli lists this value as 1138 (4th column). But a factor of two error is introduced, as the heading of the 4th column states that the diameter of Sirius is 1138 Earth *diameters*. Moving to the 5th column, $\{1138^3 = 1{,}473{,}760{,}072\}$, so Sirius is 1,473,760,072 times more voluminous than Earth.  Absent the "diameter" error, the results should be that Sirius has 569 times Earth's diameter, and 184,469,342 times its volume.

Proceeding to Table 2, a comparison of the volume of the annual orb and the volume of Sirius yields $\{1208^3/1138^3 = 1.20\}$.  Riccioli states that the annual orb is $1^2/_{15}$, or 1.13, times more voluminous than Sirius.  Absent the "diameter" error, the result should be $\{1208^3/569^3 = 9.56\}$.

The "diameter" error causes the diameter of Sirius to be too large by a factor or two, and its volume to be too large by a factor of eight.

[33] Riccioli does not include the calculations for Tables 3 and 4.  The following example is for the reader who is interested in such details, which do help to illustrate certain strengths and weaknesses of Riccioli's argument:

As an example, in the third row of Table 3, Riccioli calculates the distance to the stars based on Galileo's radius of the annual orb of 1208 Earth radii (3rd column – see previous note) if the stars are to have an annual parallax of less than 10".  This is done by considering the Earth (E), the sun (S), and a distant star (D), to form a right triangle ESD, where the sun lies at the right angle.  For an annual parallax of 10" angle EDS is 5".  Using the Galileo data, distance ES is 1208 Earth radii.  Then {ES/SD = tan EDS}, so {1208/SD = tan 5"}, and SD = 49,833,577 Earth radii. Table 3 lists this value as 49,832,416 (2nd column).  The rest of the calculations for Table 3 and 4 are done in the same way as those for Table 1 and 2 (see previous note).

The same "diameter" error that appears in Table 1 appears here, meaning Riccioli's diameter of Sirius is too large by a factor or two, and his volume too large by a factor of eight. Riccioli clearly was limited in his mathematics.  Errors appear in many places.  Of particular interest are the varying values given in Table 4.  Because of the nature of the trigonometric calculation based on 10" annual parallax, all the star distances should be a fixed multiple (approximately 41,253) of the radius of the annual orb.  Similarly, all the calculated volumes of Sirius in Table 4 should be the same multiple of the volume of the annual orb (approximately 47 given the "diameter" error, approximately 6 if done correctly).  Yet Riccioli gives values ranging from 44 to 52 – no doubt just the result of rounding error in calculations that he did multiple times, when a single calculation would have sufficed.

These errors don't alter Riccioli's argument that the stars must be huge – Sirius and Alcor still dwarf the Earth and sun, being comparable to the annual orb.  But they would have an impact on some arguments based on comparison to the annual orb that he makes shortly.

[34] See Graney, "The Telescope Against Copernicus…".

[35] In a heliocentric hypothesis obviously the Earth and the planets circled the sun.  But even in under a geo-heliocentric hypothesis such as that of Tycho Brahe, which Riccioli generally supported, the sun and moon circled the Earth but the planets circled the sun.  The idea that at least Mercury and Venus circled the sun was quite old, dating back at least to the 5th-century writer Martianus Capella – see Richard A. Jarrell, "Contemporaries of Tycho Brahe" in *The General History of*

*Astronomy: Planetary Astronomy from the Renaissance to the Rise of Astrophysics, Part A – Reni Taton, ed.* (Cambridge, 1989), p. 29.

[36] Probably a reference to Deuteronomy 4:19: "Lest perhaps lifting up thy eyes to heaven, thou see the sun and the moon, and all the stars of heaven, and being deceived by error thou adore and serve them, which the Lord thy God created for the service of all the nations, that are under heaven." I have used the Douay-Rheims version as it reflects Riccioli's language and is appropriate to the time.

[37] Riccioli also makes reference to some discussion of this by Galileo concerning Kepler, and whether Kepler understood this motion properly; see *Dialogue* pages 315-316.

[38] Today such people are usually referred to as semi-Tychonics rather than semi-Copernicans – see Christine Schofield, "The Tychonic and semi-Tychonic world systems" in *The General History of Astronomy*, p. 41.

[39] This argument is weaker than Riccioli knows, because he has made some calculation errors. In fact, his calculations should show that Sirius is 6 times more voluminous than the annual orb, not 44 (see note 57). Considering the great variability in the estimated distances here, this would not negate this part of Riccioli's argument that even using the annual orb as a standard of measure a geocentric system compares favorably to a heliocentric one, but he probably would need to reconsider his phrasing, and keep the focus on the comparison to the size of the Earth or sun, where his case is more dramatic.

[40] Riccioli only gives a short phrase, "exultasse ut gigantem ad curredam viam" (which can be compared to "exultavit ut gigans ad currendam viam suam" from the Vulgate Psalm 18:6). I have included a longer quote from the Douay-Rheims version to provide context. In modern versions of the Bible this is Psalm 19.

[41] Riccioli gives the short phrase, "vas admirabile opus excelsi" (matching the Vulgate Ecclesiasticus 43:2). English quote from the Douay-Rheims version Ecclesiasticus 43:1-3a.

[42] "Credibilior est Velocitatus Fixarum Diurna in Hypothesi Terrę quiescentis; quam Moles insana earumdem in Hypothesi Terræ per Orbem Annuum circumeuntis."